\documentclass[referee]{aa}

\usepackage{txfonts}
\usepackage{graphicx}
\usepackage{epstopdf}
\usepackage{color}
\newcommand{\pd}[2]{\frac{\partial #1}{\partial #2}}
\renewcommand{\vec}[1]{\mathbf{#1}}

\begin{document}

\title{Young stellar object jet models:\\
  From theory to synthetic observations}

\author{
  O.~Te\c{s}ileanu\inst{1}, T.~Matsakos\inst{2,3,4,9},
  S.~Massaglia\inst{5}, E.~Trussoni\inst{6}, A. Mignone\inst{5},
  N.~Vlahakis\inst{4}, K.~Tsinganos\inst{4}, M.~Stute\inst{7},\\
  V.~Cayatte\inst{8}, C.~Sauty\inst{8},
  C.~Stehl\'e\inst{9}, \and J.-P.~Chi\`eze\inst{3}
}

\authorrunning{O. Te\c{s}ileanu et al.}
\titlerunning{Synthetic observations from two-component YSO jet models}

\institute{
  ``Horia Hulubei'' National Institute of Physics and Nuclear Engineering, 30
    Reactorului Str., RO-077125 Bucure\c{s}ti-M\~agurele, Romania \and
  CEA, IRAMIS, Service Photons, Atomes et Mol\'ecules, 91191 Gif-sur-Yvette,
    France \and
  Laboratoire AIM, CEA/DSM - CNRS - Universit\'e Paris Diderot, IRFU/Service
    d'Astrophysique, CEA Saclay, Orme des Merisiers, 91191 Gif-sur-Yvette,
    France \and
  IASA \& Sect. of Astrophysics, Astronomy and Mechanics, Dept. of Physics,
    University of Athens, 15784 Zografos, Athens, Greece \and
  Dipartimento di Fisica, Universit\`a degli Studi di Torino, via Pietro Giuria
    1, 10125 Torino, Italy \and
  INAF/Osservatorio Astronomico di Torino, via Osservatorio 20, 10025 Pino
    Torinese, Italy \and
  Institute for Astronomy and Astrophysics, Section Computational Physics,
    Eberhard Karls Universit\"at T\"ubingen, Auf der Morgenstelle 10, 72076,
    T\"ubingen, Germany \and
  LUTh, Observatoire de Paris, UMR 8102 du CNRS, Universit\'e Paris Diderot,
    F-92190 Meudon, France \and
  LERMA, Observatoire de Paris, 5 Place J. Janssen, 92195 Meudon, France
}

\date{Received ?? / Accepted ??}

\abstract{
  Astronomical observations, analytical solutions and numerical simulations have
  provided the building blocks to formulate the current theory of young stellar
  object jets.
  Although each approach has made great progress independently, it is only
  during the last decade that significant efforts are being made to bring the
  separate pieces together.
}{
  Building on previous work that combined analytical solutions and numerical
  simulations, we apply a sophisticated cooling function to incorporate
  optically thin energy losses in the dynamics.
  On the one hand, this allows a self-consistent treatment of the jet evolution
  and on the other, it provides the necessary data to generate synthetic
  emission maps.
}{
  Firstly, analytical disk and stellar outflow solutions are properly combined
  to initialize numerical two-component jet models inside the computational box.
  Secondly, magneto-hydrodynamical simulations are performed in 2.5D, following
  properly the ionization and recombination of a maximum of $29$ ions.
  Finally, the outputs are post-processed to produce artificial observational
  data.
}{
  The values for the density, temperature, and velocity that the simulations
  provide along the axis are within the typical range of protostellar outflows.
  Moreover, the synthetic emission maps of the doublets [\ion{O}{i}],
  [\ion{N}{ii}] and [\ion{S}{ii}] outline a well collimated and knot-structured
  jet, which is surrounded by a less dense and slower wind, not observable in
  these lines.
  The jet is found to have a small opening angle and a radius that is also
  comparable to observations.
}{
  The first two-component jet simulations, based on analytical models, that
  include ionization and optically thin radiation losses demonstrate promising
  results for modeling specific young stellar object outflows.
  The generation of synthetic emission maps provides the link to observations,
  as well as the necessary feedback for the further improvement of the available
  models.
}


\maketitle

%%%%%%%%%%%%%%%%%%%%%%%%%%%%%%%%%%%%%%%%%%%%%%%%%%%%%%%%%%%%%%%%%%%%%%%%%%%%%
%
\section{Introduction}
%
%
%%%%%%%%%%%%%%%%%%%%%%%%%%%%%%%%%%%%%%%%%%%%%%%%%%%%%%%%%%%%%%%%%%%%%%%%%%%%%

Young stellar object (YSO) jets have been extensively studied over the past
decades, being a key element to understand the principles of star formation.
Three main approaches have been followed to address the phenomenon:
high angular resolution observations to pinpoint their properties, analytical
treatment to formulate the appropriate physical context and numerical
simulations to explore their complicated time-dependent dynamics.
Although several studies have linked any two of the above approaches, namely,
theory \& observations (e.g. Cabrit et al. \cite{Cab99}; Ferreira et al.
\cite{Fer06}; Sauty et al. \cite{Sau11}), simulations \& observations (e.g.
Massaglia et al. \cite{Mas05}; Te\c{s}ileanu et al. \cite{Tes09,Tes12}; Staff et
al. \cite{Sta10}), theory \& simulations (e.g. Gracia et al. \cite{Gra06};
Stute et al. \cite{Stu08}; \v{C}emelji\'c et al. \cite{Cem08}; Matsakos et al.
\cite{Tit08,Tit09,Tit12}; Sauty et al. \cite{Sau12}), there has not yet been
significant effort to combine all three of them.

The basic YSO jet properties have been well known since more than a decade.
They are accretion powered (e.g. Cabrit et al. \cite{Cab90}; Hartigan et al.
\cite{Har95}), they propagate for thousands of $\mathrm{AU}$ (e.g. Hartigan et
al. \cite{Har04}) and have a velocity of a few hundreds $\mathrm{km\,s}^{-1}$
(e.g. Eisl\"offel \& Mundt \cite{Eis92}; Eisl\"offel et al. \cite{Eis94}).
Such outflows are well collimated (e.g. Ray et al. \cite{Ray96}) with a jet
radius of $50\,\mathrm{AU}$ (e.g. Dougados et al. \cite{Dou00}) and a structure
that consists of several knots (termed HH objects) which are shocks occuring
from speed variabilities (see Reipurth \& Bally \cite{Rei01} for a review).
YSO jets, among other mechanisms, are capable of removing a significant amount
of angular momentum both from the disk and the star.
The former is crucial for accretion (e.g. Lynden-Bell \& Pringle \cite{Lyn74})
and the latter for the protostellar spin-down (e.g. Sauty et al. \cite{Sau11};
Matt et al. \cite{Mat12}).
These two processes are necessary to allow the star to enter the main sequence.
Moreover, jet ejection takes place close to the central object (e.g. Ray et al.
\cite{Ray07}, and references therein) with evidence of being either of a disk or
a stellar origin, or a combination of the two (e.g. Edwards et al. \cite{Edw06};
Kwan et al. \cite{Kwa07}).

Numerical simulations have studied in detail the launching, collimation, and
propagation mechanisms of YSO outflows adopting two main approaches.
The first one assumes that the ejection takes place below the computational box,
and hence the flow properties are specified as boundary conditions on the lower
ghost zones.
This approach is appropriate to model the large scale jet structure and allows
to explore a wide parameter space (e.g. Ouyed \& Pudritz
\cite{Ouy97a,Ouy97b,Ouy99}; Krasnopolsky et al. \cite{Kra99}; Anderson et al.
\cite{And05}; Pudritz et al. \cite{Pud06}; Fendt \cite{Fen06,Fen09}; Matt \&
Pudritz \cite {Mat08}; Matsakos et al. \cite{Tit08,Tit09,Tit12}; Staff et al.
\cite{Sta10}; Te\c{s}ileanu et al. \cite{Tes12}).
The second approach evolves the flow together with the dynamics of the disk, and
thus is very demanding computationally.
Even though it cannot follow the flow at large scales, it does provide a
self-consistent treatment of the YSO-jet system linking mass loading with
accretion (e.g. {Casse \& Keppens \cite{Cas02,Cas04}; Meliani et al.
\cite{Mel06}; Zanni et al. \cite{Zan07}; Tzeferacos et al. \cite{Tze09,Tze13};
Sheikhnezami et al. \cite{She12}; Fendt \& Sheikhnezami \cite{Fen13}).
These two approaches are not competing, on the contrary, they are both necessary
and complementary to each other for the study of jets.
Here, we adopt the former because we focus on the propagation scales.

From the theoretical point of view, Vlahakis \& Tsinganos (\cite{Vla98}) have
derived the possible classes of analytical steady-state and axisymmetric
magneto-hydrodynamical (MHD) outflow solutions based on the assumption of radial
or meridional self-similarity.
Two families have emerged, each one being appropriate to describe disk or
stellar jets.
In the following, we refer to them as Analytical Disk Outflows (ADO) and
Analytical Stellar Outflows (ASO), respectively.
In fact, the Parker's solar wind (\cite{Par58}), the Blandford \& Payne model
(\cite{Bla82}) as well as other previously known solutions (e.g. Sauty \&
Tsinganos \cite{Sau94}; Trussoni et al. \cite{Tru97}) were all found to be
specific cases within that framework.

One implication of self-similarity is that the shape of the critical surfaces of
the flow is either conical or spherical, for the ADO or ASO solutions,
respectively.
For instance, for the ADO models, this reflects the assumption that the outflow
variables have a scaling such that the launching mechanism does not depend on
any specific radius of a Keplerian disk.
If the flow quantities are known for one fieldline, the whole solution can be
reconstructed.
Within this context, self-collimated outflows can be derived and appropriately
parameterized to match most of the observed properties, such as velocity and
density profiles (see Tsinganos \cite{Tsi07} for a review).
However, a physically consistent treatment of the energy equation cannot be
easily incorporated in self-similar models.
Consequently, either a polytropic flow is assumed, or the heating/cooling source
term is derived a posteriori.
Moreover, the symmetry assumptions make the ADO solutions diverge on the axis,
whereas the ASO models become inappropriate to describe disk winds.
Nevertheless, this makes the two families of solutions complementary to each
other, with their numerical combination naturally addressing their shortcomings.

Various physical and numerical aspects of each of the ADO and ASO solutions have
been studied separately (Gracia \cite{Gra06}; Matsakos et al. \cite{Tit08};
Stute et al. \cite{Stu08}) proving the robustness of their stability.
Subsequently, several two-component jet models have been constructed and
simulated examining the parameter space of their combination.
Velocity variabilities were also included and were found to produce knots that
resemble real jet structures (Matsakos et al. \cite{Tit09,Tit12}).
On the other hand, Te\c{s}ileanu et al. (\cite{Tes09,Tes12}) specified typical
YSO flow variables on the lower boundary of the computational box and applied
optically thin radiation losses during the numerical evolution.
They studied shocks in the presence of a realistic cooling function, and also
created emission maps and line ratios, directly comparing with observations.

The present work combines both of the above approaches.
It initializes a steady-state two-component jet throughout the computational
domain (from Matsakos et al. \cite{Tit12}) while imposing a physically
consistent treatment of the energy equation (as in Te\c{s}ileanu et al.
\cite{Tes12}).
Apart from the calculation of optically thin radiation losses, the coevolution
of the ionization allows also the generation of synthetic emission maps
that we discuss in the context of typical YSO jets.
Stute et al. (\cite{Stu10}) were the first to compare numerically modified
analytical models to observed jets.
They simulated truncated ADO solutions and then post-processed the outputs to
calculate the emission.
Comparison with observed jet radii provided a good match for the several cases
they examined.

The structure of this paper is as follows.
Section \S\ref{sec:setup} presents the theoretical framework and describes the
technical part of the numerical setup.
Section \S\ref{sec:results} discusses the results 
and Sect.~\S\ref{sec:conclusions} reports our conclusions.

%%%%%%%%%%%%%%%%%%%%%%%%%%%%%%%%%%%%%%%%%%%%%%%%%%%%%%%%%%%%%%%%%%%%%%%%%%%%%
%
\section{Setup}
  \label{sec:setup}
%
%
%%%%%%%%%%%%%%%%%%%%%%%%%%%%%%%%%%%%%%%%%%%%%%%%%%%%%%%%%%%%%%%%%%%%%%%%%%%%%

%%%%%%%%%%%%%%%%%%%%%%%%%%%%%%%%%%%%%%%%%%%%%%%%%%%%%%%%%%%%%%%%%%%%%%%%%%%%%
\subsection{MHD equations}

The ideal MHD equations written in quasi-linear form are:
\begin{equation}
  \frac{\partial\rho}{\partial t} + \nabla \cdot (\rho \vec V) = 0\,,
\end{equation}
\begin{equation}
  \frac{\partial\vec V}{\partial t} + (\vec V \cdot \nabla)\vec V
    + \frac{1}{\rho}\vec B \times(\nabla\times\vec B)
    + \frac{1}{\rho}\nabla P = - \nabla \Phi\,,
\end{equation}
\begin{equation}\label{eqEnerg}
  \frac{\partial P}{\partial t} + \vec V \cdot \nabla P
    + \Gamma P \nabla \cdot \vec V = \Lambda\,,
\end{equation}
\begin{equation}
  \frac{\partial\vec B}{\partial t} + \nabla \times (\vec B \times \vec V)
    = 0\,,
\end{equation}
where $\rho$, $P$, $\vec V$, and $\vec B$ denote the density, gas pressure,
velocity, and magnetic field, respectively.
The factor $(4\pi)^{-1/2}$ is absorbed in the definition of $\vec B$, which of
course is divergence free, $\nabla\cdot\vec B = 0$.
The gravitational potential is given by $\Phi = -GM/R$, where $G$ the
gravitational constant, $M$ the mass of the central object and $R$ the spherical
radius.
Finally, $\Gamma = 5/3$ is the ratio of the specific heats and $\Lambda$
represents optically thin radiation losses, which are presented in detail in
Sect.~\ref{sec:cooling}.
Even though the simulations are performed in code units, our results are
presented directly in physical units.

%%%%%%%%%%%%%%%%%%%%%%%%%%%%%%%%%%%%%%%%%%%%%%%%%%%%%%%%%%%%%%%%%%%%%%%%%%%%%
\subsection{Numerical models}

We take several steps to set up the initial conditions of the two-component jet
simulations.
In summary, analytical MHD outflow solutions are employed, normalized to each
other and then properly combined.
A low resolution simulation is carried out and the final steady state is saved.
The output data are used to initialize our main simulations, in which we apply a
velocity variability at the base, optically thin radiation losses and Adaptive
Mesh Refinement (AMR) that significantly increases the effective resolution.

The ADO solution is adopted from Vlahakis et al. (\cite{Vla00}) and describes a
magneto-centrifugally accelerated disk wind that successfully crosses all three
magnetosonic critical surfaces.
The ASO model is taken from Sauty et al. (\cite{Sau02}) and is a pressure driven
solution with a large lever arm capable of spinning down the protostar (see
Matsakos et al. \cite{Tit09} for more details on the implementation of the
solutions).

The two-component jets are initialized with the stellar outflow replacing the
inner regions of the disk wind.
The normalization and combination of the solutions is based on the following
numerical and physical arguments, an approach adopted from Matsakos et al.
(\cite{Tit12}).
First, the solutions are scaled in order to correspond to the same central mass.
Then, a matching surface is chosen at an appropriate location such that the
shape of the magnetic field of the disk wind matches approximately the geometry
of the ASO fieldlines.
Finally, we require that the magnitude of $\vec B$, which is provided by each of
the analytical models on that surface, has a similar value.

The transition between the two solutions is based on the magnetic flux function
$A$, which essentially labels the fieldlines of each solution (see Vlahakis et
al. \cite{Vla00} and Sauty et al. \cite{Sau02}).
Initially, we create the variable
$A_1 = A_\mathrm{D} + A_\mathrm{S} + A_\mathrm{c}$, simply by adding the flux
functions of the ADO and ASO components, with $A_\mathrm{c}$ a normalization
constant.
We then create the variable $A_2$ by exponentially smoothing out the solutions
around the matching surface, $A_\mathrm{mix}$, such that the stellar solution
dominates close to the axis and the disk wind at the outer regions, i.e.
\begin{equation}
  A_2 = \left\{1-\exp
    \left[-\left(\frac{A_1}{A_\mathrm{mix}}\right)^2\right]\right\}
    A_\mathrm{D} + \exp
    \left[-\left(\frac{A_1}{A_\mathrm{mix}}\right)^2 \right]
    A_\mathrm{S}\,.
\end{equation}
This new magnetic flux function, $A_2$, is used to initialize most of the
physical quantities $U$, namely $\rho$, $P$, $\vec V$, $A$ and $B_\phi$, with
the help of the following formula:
\begin{equation}
  U = \left\{1-\exp
    \left[-\left(\frac{A_2}{A_\mathrm{mix}}\right)^2\right]\right\}
    U_\mathrm{D} + \exp
    \left[-\left(\frac{A_2}{A_\mathrm{mix}}\right)^2 \right]
    U_\mathrm{S}\,.
\end{equation}
Finally, the poloidal magnetic field is derived from the magnetic flux function,
i.e. $\vec{B}_\mathrm{p} = (\nabla A\times\hat\phi)/r$, and hence it
divergent-free by definition.

At this stage, the two-component jet model is evolved adiabatically in time
until it reaches a steady state.
For efficiency, we perform the simulation on a low resolution grid, i.e. $128$
zones in the radial direction and $1024$ in the vertical.
We point out that the level of refinement does not affect the final outcome of
the simulation.
On the contrary, the final configuration is a well maintained steady state and
it is obtained independently of the resolution or the mixing parameters, see
Matsakos et al. (\cite{Tit09}).
The use of AMR is required for the correct treatment of the cooling, as
explained later, and does not affect the steady state of the adiabatic jet
simulation.

This steady state is then used as initial conditions for the simulation that
includes optically thin radiation cooling, also imposing fluctuations in the
flow.
The variability in the velocity is achieved by multiplying its longitudinal
component with a sinusoidal dependence in time and gaussian in space, i.e.
$V_z \to f_\mathrm{S}V_z$, with
\begin{equation}
  f_\mathrm{S}(r, t) =
    1 + p\exp\left[-\left(\frac{r}{r_\mathrm{var}}\right)^2\right]
    \sin\left(\frac{2\pi t}{t_\mathrm{var}}\right)\,,
\end{equation}
where $p = 20\%$, $r$ is the cylindrical radius,
$r_\mathrm{var} = 50\,\mathrm{AU}$, and $t_\mathrm{var} = 3.7\,\mathrm{yr}$.
The variability of the velocity is chosen small enough such that we can assume
a constant pre-ionization of the shocked material (Cox \& Raymond \cite{Cox85}).

%%%%%%%%%%%%%%%%%%%%%%%%%%%%%%%%%%%%%%%%%%%%%%%%%%%%%%%%%%%%%%%%%%%%%%%%%%%%%
\subsection{Cooling function}
  \label{sec:cooling}

We take advantage of the Multi-Ion Non-Equilibrium (MINEq) cooling module
developed by Te\c{s}ileanu et al. (\cite{Tes08}), that includes an ionization
network and a 5-level atom model for radiative transitions.
Previous approaches followed the evolution in time of only the ionization of 
hydrogen, assuming equilibrium ionization states for the other elements of
interest.
In the context of strong shocks propagating and heating the plasma, as
encountered in variable stellar jets, the assumption of equilibrium limited the
accuracy and reliability of the simulation results. 

The MINEq cooling function includes a network of a maximum of $29$ ion species, 
selected appropriately to capture most of the radiative losses in 
YSO jets up to temperatures of $200\,000\,\mathrm{K}$.
These ion species represent the first five ionization states (from I to V) of C,
N, O, Ne, and S, as well as \ion{H}{i}, \ion{H}{ii}, \ion{He}{i}, and
\ion{He}{ii}. 
In more recent versions, the cooling function implementation allows the user to
select the number of ionization states required for each simulation 
 (depending on shock strength, some of the upper-lying ionization states may
be disabled).
In the present work, the first three ionization states were used, a choice that
is adequate for the physical conditions developing in the areas of interest.

For each ion, one additional equation must be solved:
\begin{equation}\label{eq:species}
  \pd{(\rho X_{\kappa,i})}{t} + \nabla\cdot\left(\rho X_{\kappa,i}\vec{v}\right)
    = \rho S_{\kappa,i}\,,
\end{equation}
coupled to the original system of MHD equations.
In the equation above, the first index ($\kappa$) describes the element, while
the second index ($i$) corresponds to the ionization state.
Specifically, $X_{\kappa,i} \equiv N_{\kappa,i} / N_{\kappa}$ is the ion number
fraction, $N_{\kappa,i}$ is the number density of the $i$-th ion of element
$\kappa$, and $N_{\kappa}$ is the element number density.
The source term $S_{\kappa,i}$ accounts for the effect of the ionization and 
recombination processes.

For each ion, the collisionally excited line radiation is computed and the total
line emission from these species enters in the source term $\Lambda$ of
Eq.~(\ref{eqEnerg}).
This provides an adequate approximation of radiative cooling for the conditions
encountered in YSO jets.
Depending on the conditions, more ion species may be included.

Cooling introduces an additional timescale in MHD numerical simulations together
with the dynamical one.
In regions where sudden compressions/heating of the gas occur, the timescale for
cooling and ionization of the gas becomes much smaller and hence dominates the
simulation timestep.
Consequently, a larger number of integration steps are needed and the total
duration of the simulations increases considerably.

Finally, another aspect to consider is the numerical resolution.
This is required for a sufficiently accurate and reliable physical description
of the processes occuring in the post-shock zone behind a shock front,
especially concerning the ionization state of the plasma.
As discussed in previous work, a resolution higher than $10^{12}\,\mathrm{cm}$ 
($\sim$$0.07$AU) per integration cell is needed in order to adequately resolve 
the physical parameters after the shock front (Mignone et al. \cite{Min09}).
Therefore, Adaptive Mesh Refinement is necessary in order to treat correctly the
dynamics/cooling while retaining high efficiency for the simulation.

%%%%%%%%%%%%%%%%%%%%%%%%%%%%%%%%%%%%%%%%%%%%%%%%%%%%%%%%%%%%%%%%%%%%%%%%%%%%%
\subsection{Numerical setup}

We use PLUTO, a numerical code for computational astrophysics\footnote{Freely
available at \texttt{http://plutocode.ph.unito.it}}, to carry out the MHD
simulations (Mignone et al. \cite{Mig07}).
We choose cylindrical coordinates $(r,z)$ in 2.5D, assuming axisymmetry to
supress the third dimension.
Integration is performed with the HLL solver with second order accuracy in both
space and time, whereas the condition $\nabla\cdot\vec{B}$ = 0 is ensured with
hyperbolic divergence cleaning.
Since the jet kinematics involve length and time scales much larger that those
controlled by optically thin radiation losses (e.g. the post-shock regions), 
Adaptive Mesh Refinement (AMR) is adopted (Mignone et al. \cite{Mig12}).
The refinement strategy is based on the second derivative error norm (see
Mignone et al. \cite{Mig12}) taken for the quantity defined by the product of
the temperature with radius.
Such a criterion is appropriate to resolve the shocks as well as the region
around the axis.

We perform two simulations, one using the detailed MINEq cooling function, and
another that only evolves the ionization of H (SNEq - Simplified Non-Equilibrium
cooling).
Our computational box spans $[0,\, 80]\,\mathrm{AU}$ in the radial direction and
$[100,\, 740]\,\mathrm{AU}$ in the longitudinal.
For the model employing SNEq, the base grid is $64\times512$ with $5$ levels of
refinement that provides an effective resolution of $16384$ zones along the
axis, equivalent to one cell for each $0.04\,\mathrm{AU}$.
For the more computationally expensive cooling module MINEq, our base grid is
$32\times256$ with also $5$ levels of refinement, equivalent to one cell per
$0.08\,\mathrm{AU}$.
We prescribe outflow conditions on the top and right boundaries, axisymmery on
the left and we keep fixed all quantities to their initial values on the bottom
boundary.
We note that the two-component jet model is initialized everywhere inside the
computational box and hence we do not model the bow shock or the acceleration
regions, but rather a part of the outflow in between.
Ionization is initialized throughout the entire integration domain with
equilibrium values resulting from the local physical conditions (temperature and
density).

Finally, a comment on the location of the lower edge of the computational box.
Apart from the fact that the paper focuses and attempts to address the
propagation scales, the choice for the height of the bottom boundary is based on
the following two reasons.
Firstly, the high resolution required to treat properly the launching regions of
the stellar and disk outflows is a prohibitive to be combined together with
distances on the order of hundreds of AU.
Secondly, the source term in the energy equation close to the base of the jet
is complex and not well-known.
The mass loading of the field lines, and in some cases the acceleration,
requires some sort of heating which involves extra assumptions and additional
parameters.
Instead, we have decided to take advantage of the available analytical solutions
and start the computation at a large distance where such heating terms are not
present.

%%%%%%%%%%%%%%%%%%%%%%%%%%%%%%%%%%%%%%%%%%%%%%%%%%%%%%%%%%%%%%%%%%%%%%%%%%%%%
\subsection{Post-processing and emission maps}

The code provides as output the maps of all physical quantities at specified
intervals of time.
Further treatment of the data is required before they can be directly compared
to observations.

The first post-processing routine is the calculation of 2D emission maps from
the simulation output.
The collisionally excited emission lines of observational interest treated in the present work
are the forbidden emission line doublets of \ion{N}{ii} (6584 + 6548 \AA),
\ion{O}{i} (6300 + 6363 \AA), and \ion{S}{ii} (6717 + 6727 \AA).
Next, the 3D integration of this axially-symmetric output is performed, to
account for the particular geometry encountered for each source.
At this stage, a tilt angle of the simulated jet with respect to the line of
sight may be applied.
Then, the 3D object is projected on a surface perpendicular to the line of sight
(equivalent to the ``plane of sky''), and the distance to the object is taken
into consideration in order to convert all data to the same units as
observations.
Note that a Point Spread Function (PSF) is also applied in the simplified form
of a Gaussian function with a user-defined width, in order to simulate the
effect of the observing instrument.\footnote{
  All post-processing routines are included in user-friendly templates in the
  current distribution of the PLUTO code.
}

%%%%%%%%%%%%%%%%%%%%%%%%%%%%%%%%%%%%%%%%%%%%%%%%%%%%%%%%%%%%%%%%%%%%%%%%%%%%%
%
\section{Results}
  \label{sec:results}
%
%
%%%%%%%%%%%%%%%%%%%%%%%%%%%%%%%%%%%%%%%%%%%%%%%%%%%%%%%%%%%%%%%%%%%%%%%%%%%%%

%%%%%%%%%%%%%%%%%%%%%%%%%%%%%%%%%%%%%%%%%%%%%%%%%%%%%%%%%%%%%%%%%%%%%%%%%%%%%
\subsection{Dynamics}
  \label{sec:dynamics}

\begin{figure}
\centering
  \resizebox{\hsize}{!}{\includegraphics{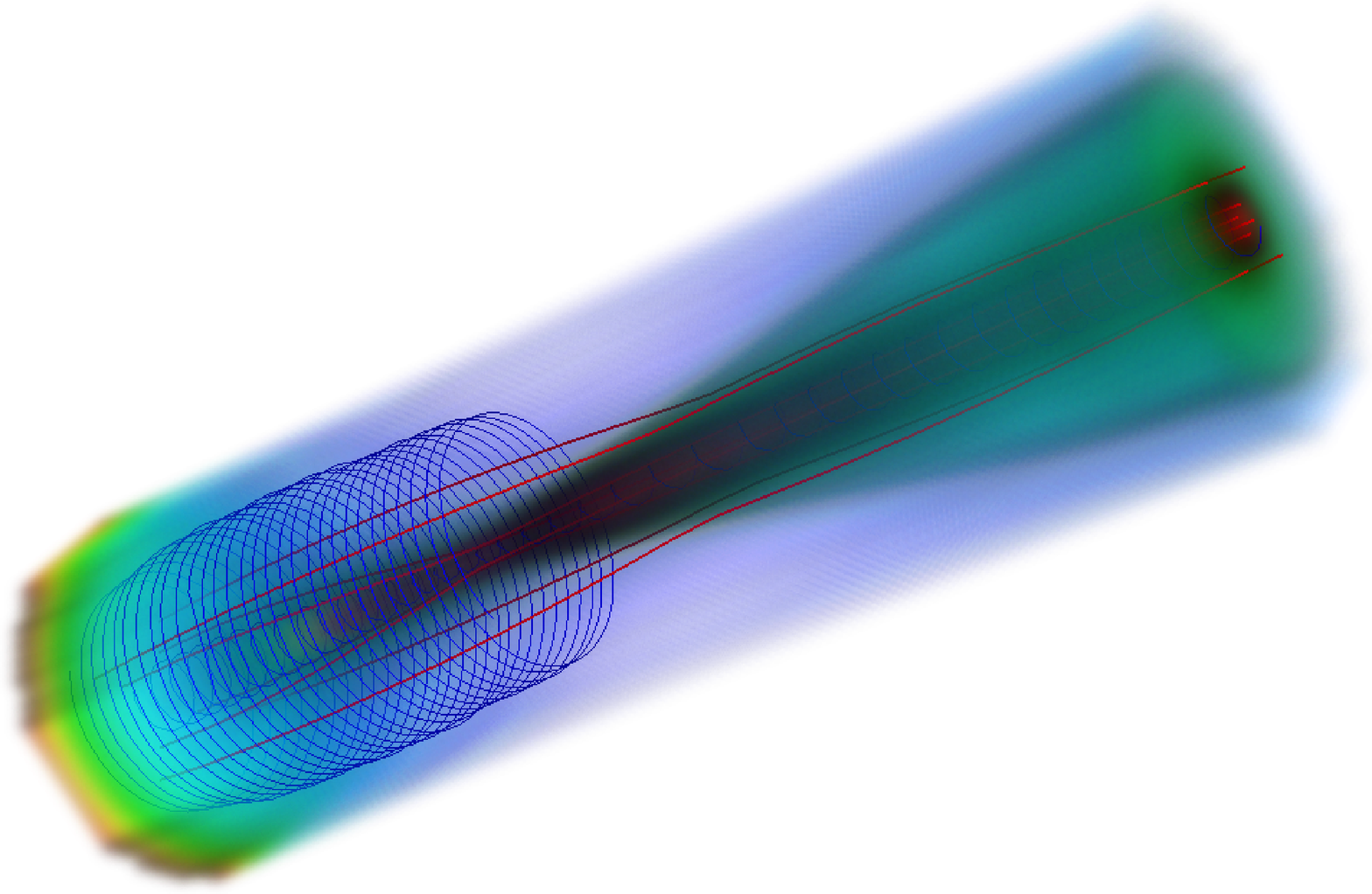}}
  \resizebox{\hsize}{!}{\includegraphics{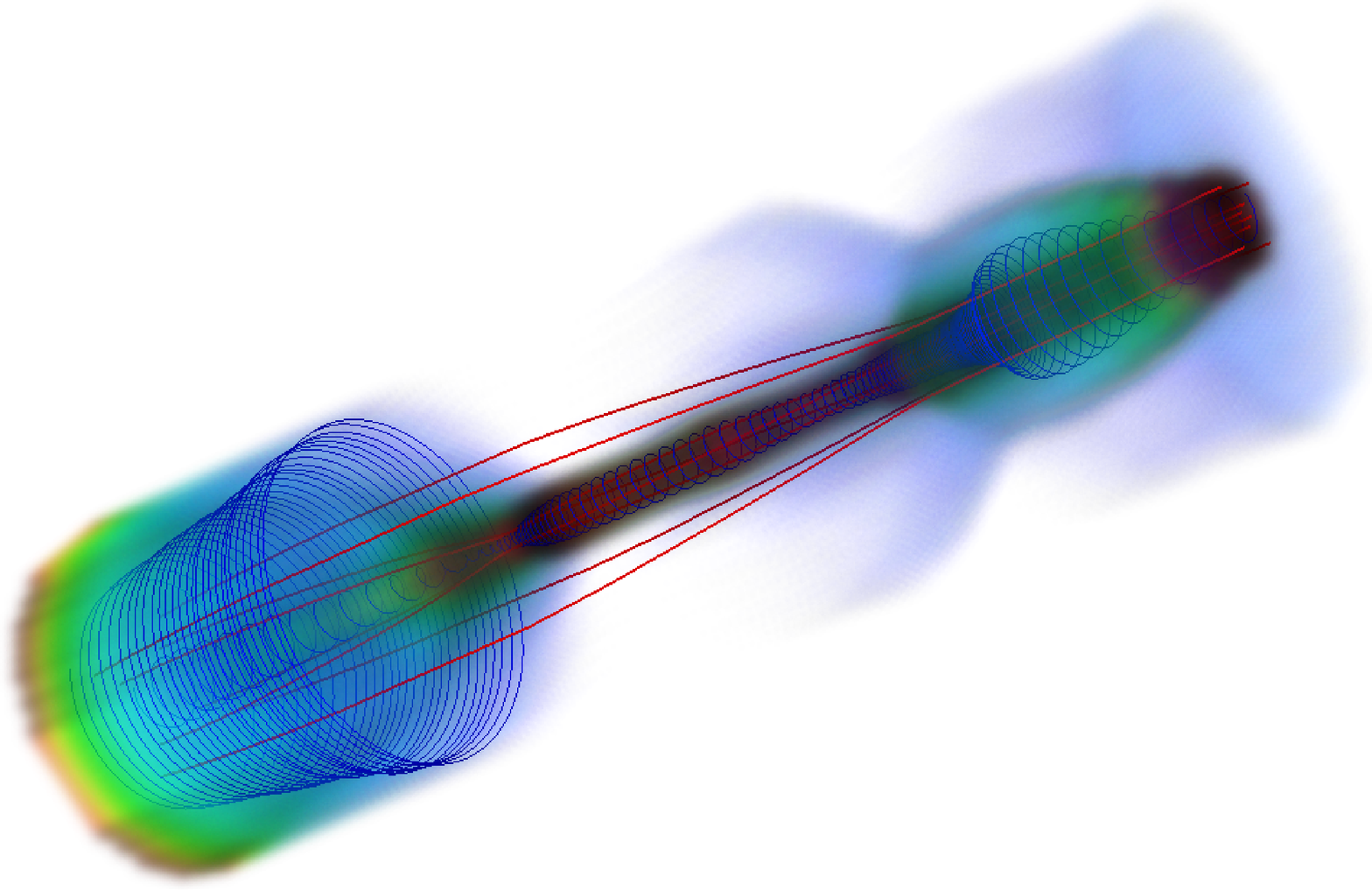}}
  \caption{3D representation of the density distribution of the initial
    conditions (top) and at a later time during evolution (bottom).
    Blue/red corresponds to low/high density, the thin blue lines denote the
    magnetic field and the thick red lines the flow.}
  \label{fig:dynamics}
\end{figure}
The initial conditions correspond to a steady-state magnetized self-collimated
jet, a configuration that is reached after the combination and adiabatic
evolution of the two wind components, see top panel of Fig.~\ref{fig:dynamics}.
Its inner part represents a hot stellar outflow which is collimated by the hoop
stress provided by the magnetic field of the surrounding disk wind.
The longitudinal velocity decreases with radius, i.e. the flow consists of a fast
jet close to the axis and a slow wind at the outer radii.
During the first steps of the simulation the cooling function lowers the
temperature of the plasma, especially in the inner hot regions.
Due to the pressure drop, the collimating forces squeeze the jet and in turn the
jet radius is decreased, see bottom panel of Fig.~\ref{fig:dynamics} (and
Sect.~\ref{sec:jetradius} for a discussion).
Moreover, the variability applied on the bottom boundary produces shocks that
propagate along the axis and lead to the formation of a knot-structured jet.
These internal shocks heat locally the gas and in turn the post-shock regions
are susceptible to larger energy losses and strong emission.

\begin{figure}
\centering
  \resizebox{\hsize}{!}{\includegraphics{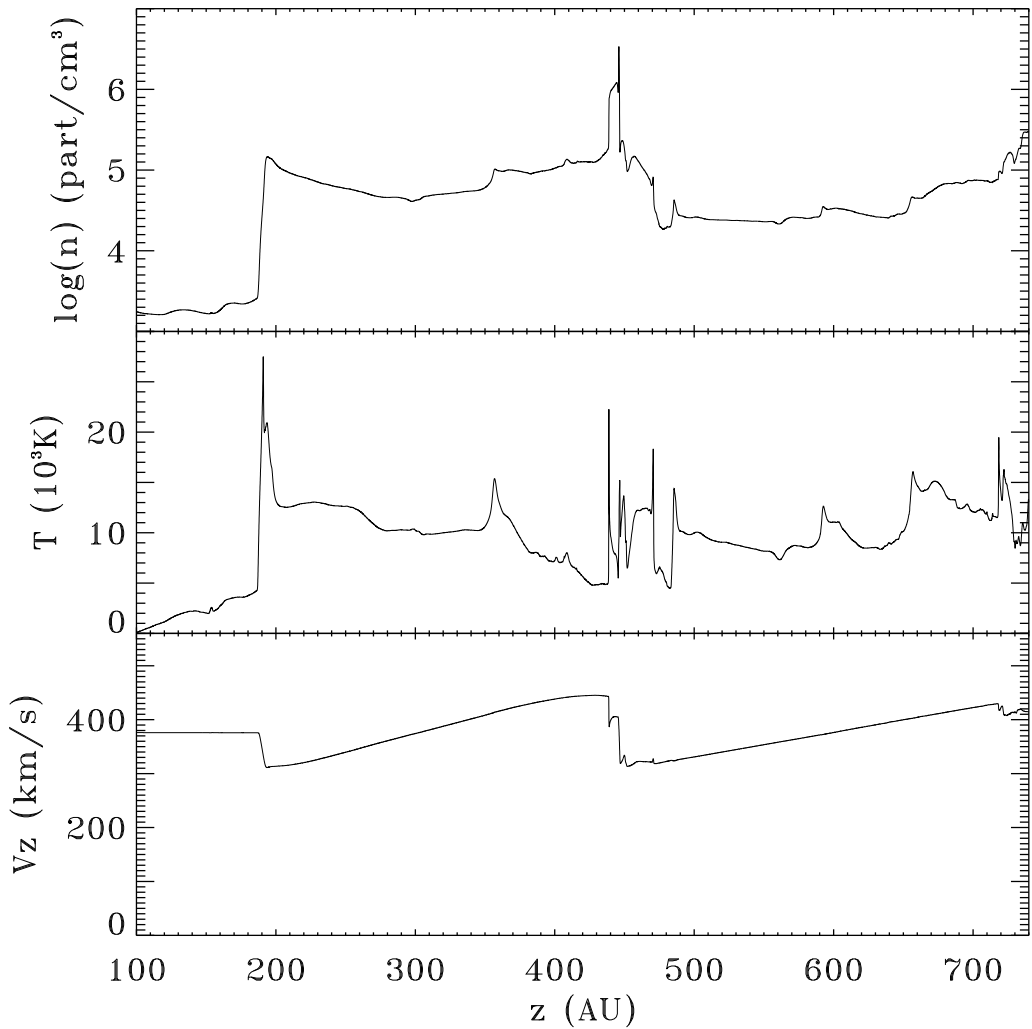}}
  \caption{Logarithmic number density (top), temperature (middle) and velocity
    (bottom) profiles along the axis, for the model adopting SNEq.
    The jet is moving from left to right and the displayed moment corresponds to
    $t = 12\,\mathrm{yr}$.}
  \label{fig:profiles_vs_z_sneq}
\end{figure}
\begin{figure}
\centering
  \resizebox{\hsize}{!}{\includegraphics{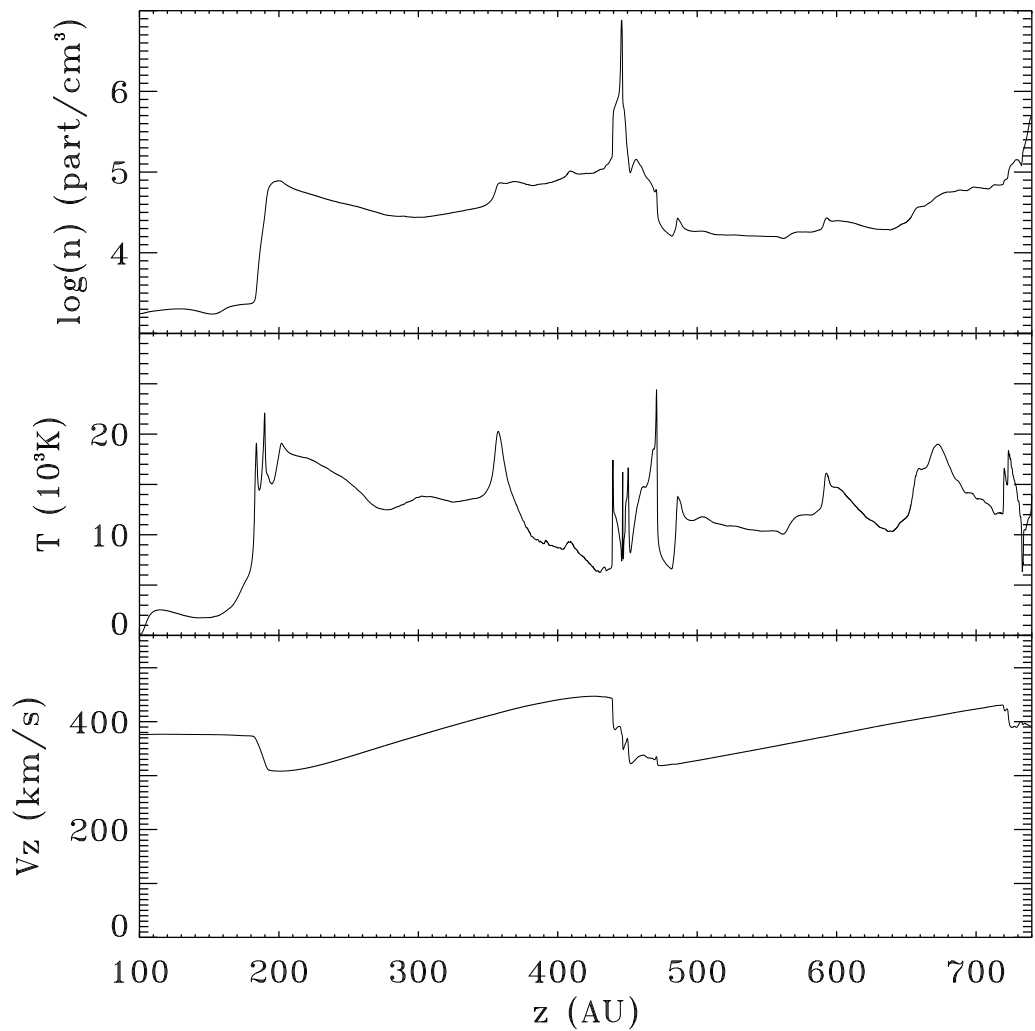}}
  \caption{Logarithmic density (top), temperature (middle) and velocity (bottom)
    profiles along the axis, for the model adopting MINEq.
    The instance corresponds to $t = 12\,\mathrm{yr}$.}
  \label{fig:profiles_vs_z_mineq}
\end{figure}
Figure~\ref{fig:profiles_vs_z_sneq} shows the average values of the number
density, temperature, and velocity of the outflow around the axis (for
$r \leq 0.5\,\mathrm{AU}$), for the simulation performed with the SNEq module.
Figure~\ref{fig:profiles_vs_z_mineq} displays the same quantities for the model
which adopts the MINEq cooling function.
The number density is on the order of $10^4-10^5\,\mathrm{cm}^{-3}$, with higher
local values at overdensities produced by the variability of the flow.
A strong shock can be observed approximately at $450\,\mathrm{AU}$, in
accordance with the negative slope of the velocity profile.
The jet temperatute is on the order of $10\,000\,\mathrm{K}$, but may exceed
$20\,000\,\mathrm{K}$ at the post-shock regions.
The jet velocity is on average $380\,\mathrm{km\,s}^{-1}$ whereas the speed of
the shocks relative to the bulk flow is $\sim$$100\,\mathrm{km\,s}^{-1}$.
In general, both models capture correctly the typical values of observed YSO
jets.

Even though the two simulations incorporate different cooling functions, the
dynamics between the two simulations are roughly similar.
This suggests that the simpler SNEq module is a good approximation to treat the
energy losses during the temporal evolution.
For this reason, this module is employed in a future work that studies the full
3D jet structure (Matsakos et al. in preparation).
However, the aim of the present paper is the calculation of line emissions.
Therefore in the following we focus on the simulation that adopts the MINEq
cooling function.

\begin{figure}
\centering
  \resizebox{0.85\hsize}{!}{\includegraphics{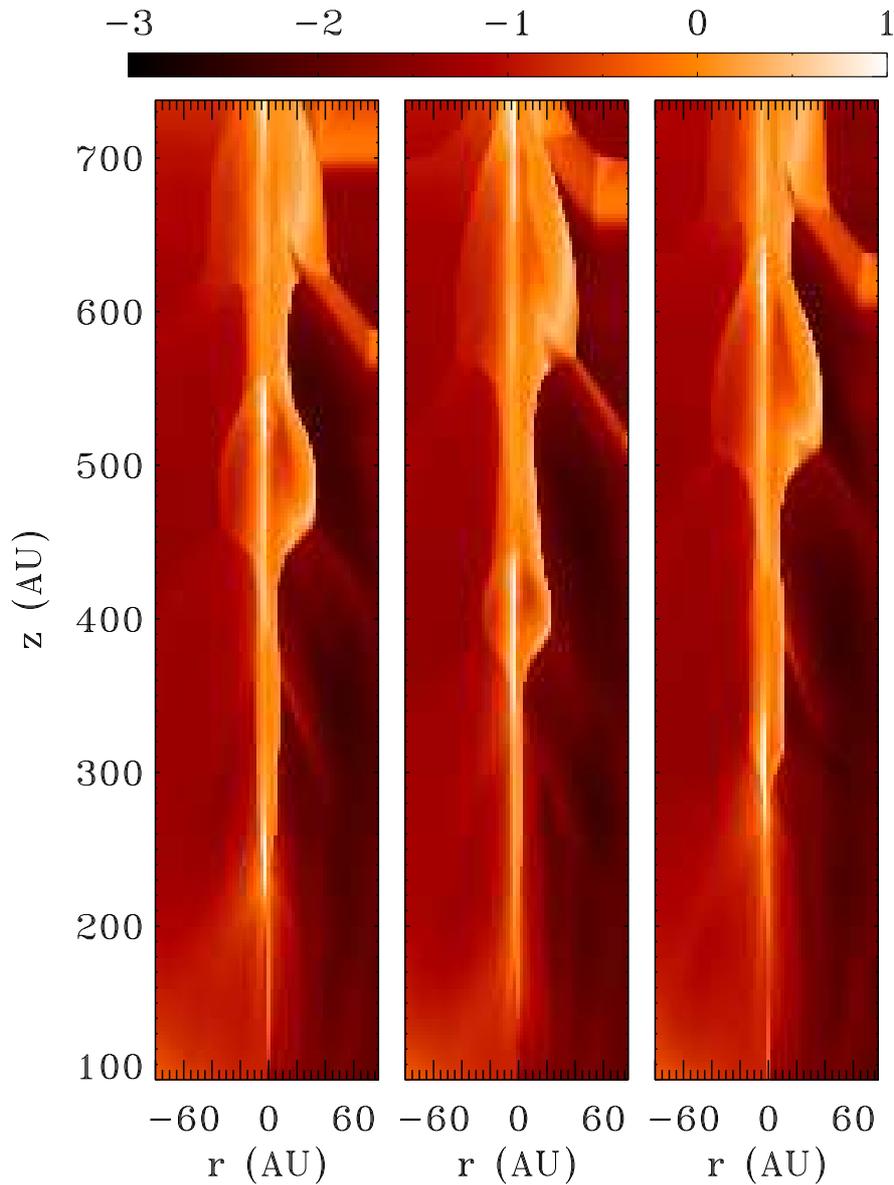}}
  \caption{Logarithmic maps of the density in normalized units of
    $10^4\,\mathrm{cm}^{-3}$ (left hand side of each
    panel), and the temperature in $10^4\,\mathrm{K}$ (right hand side of each
    panel), for the model employing the MINEq cooling.
    From left to right, three moments of the temporal evolution are shown, i.e.
    $16.8$, $19.2$, and $21.6$ years.
    The jet is propagating upwards, and specifically, the high density region
    located between $200$ and $300\,\mathrm{AU}$ in the left panel, is found
    between $400$ and $500\,\mathrm{AU}$ in the middle, and between $600$ and
    $700\,\mathrm{AU}$ in the right panel.}
  \label{fig:dn_tmp_mineq}
\end{figure}
Figure~\ref{fig:dn_tmp_mineq} shows the 2D logarithmic distribution of the
density and temperature at three evolutionary moments separated by $2.4$ years.
The higher density jet material defines the positions of the propagating knots.
For instance, a knot that can be seen at $\sim$$250\,\mathrm{AU}$ in the left
panel, is located at $\sim$$450\,\mathrm{AU}$ in the central panel, and at
$\sim$$650\,\mathrm{AU}$ in the right one.
The surface close to the axis that separates the inner hot outflow from the
outer cooler gas, is a weak oblique shock.
It forms due to the recollimation of the flow and disconnects causally the jet
from its launching region since no information can propagate backwards (Matsakos
et al. \cite{Tit08,Tit09}).

\begin{figure}
\centering
  \resizebox{\hsize}{!}{\includegraphics{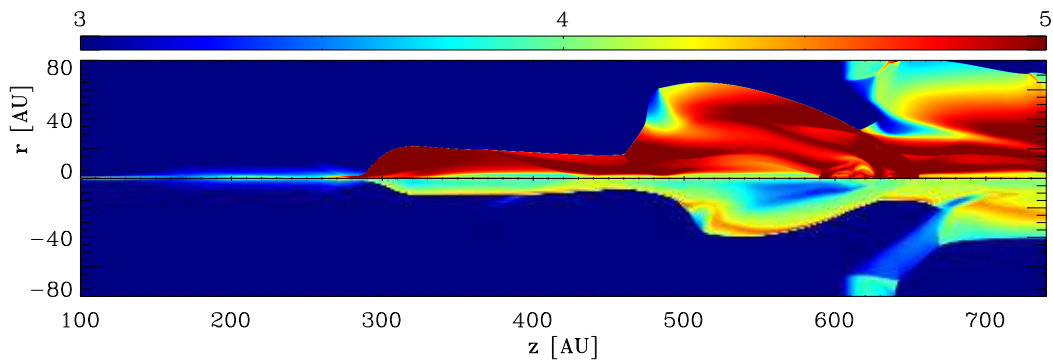}}
  \caption{Comparison of the logarithmic temperature distribution and jet radius
    for an adiabatic evolution (top) and with the MINEq cooling module
    (bottom).
    The snapshot corresponds to $t = 21.6$.}
  \label{fig:adiabatic}
\end{figure}
In order to estimate the effects of the energy losses in the dynamics of the
jet, we plot in Fig.~\ref{fig:adiabatic} the temperature distribution of an
adiabatic evolution (top) together with that of the simulation adopting MINEq
(bottom).
Optically thin radiation reduces the pressure of the inner hot flow and thus the
radius of the emitting jet is found by a factor of $2$ smaller than the
adiabatic one due to the unbalanced collimating forces.
On the other hand, note that the predicted temperature of the adiabatic model
can be by an order of magnitude higher which in turn may lead to the
overestimation of the emission.

Since the final configuration of the adiabatic simulation shows a different
morphology than the one with the imposed cooling, we should examine for
consistency the bottom boundary conditions.
Namely, we should check whether the quantities imposed at the lower edge of the
MINEq simulation, that are given by the adiabatic model, affect or enforce
anyhow the evolution of the system.
However, the effects of the optically thin radiation cooling are significant
mainly where the outflow is hot, i.e. close to the axis
(Fig.~\ref{fig:adiabatic}).
In turn, both at the base of the computational box as well as at outer radii,
the adiabatic evolution is very similar to that with cooling.
Therefore, we argue that the bottom boundary conditions are consistent with the
MINEq simulation.

%%%%%%%%%%%%%%%%%%%%%%%%%%%%%%%%%%%%%%%%%%%%%%%%%%%%%%%%%%%%%%%%%%%%%%%%%%%%%
\subsection{Ionization}

\begin{figure}
\centering
  \resizebox{0.85\hsize}{!}{\includegraphics{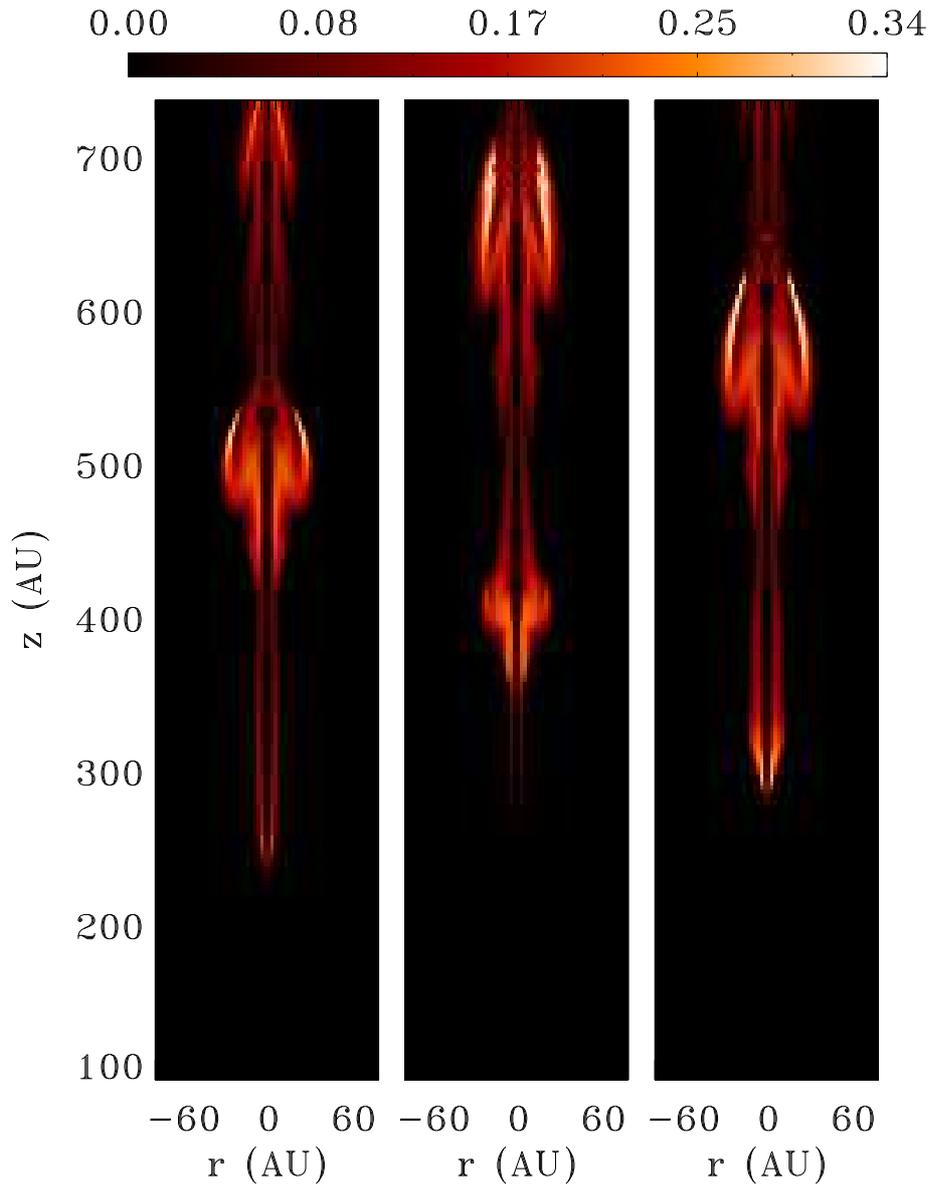}}
  \caption{Maps of the total ionization fraction of the jet material for the
    three moments in the evolution shown in Fig.~\ref{fig:dn_tmp_mineq}.}
  \label{fig:ion_evol_mineq}
\end{figure}
In the analytical models employed here, no pre-existing condition of ionization
of the jet material was set.
Ionization is self-consistently computed during the evolution, snapshots of
which are shown in Fig.~\ref{fig:ion_evol_mineq}.
The flow is locally ionized when the temperature rises, and as a result, the jet
ionization is very low close to the origin and gets higher within the knots
($\sim$$34\%$).
Note that the degree of pre-ionization may affect the line emission as discussed
in Te\c{s}ileanu et al. (\cite{Tes12}).

%%%%%%%%%%%%%%%%%%%%%%%%%%%%%%%%%%%%%%%%%%%%%%%%%%%%%%%%%%%%%%%%%%%%%%%%%%%%%
\subsection{Emission maps}

\begin{figure}
\centering
  \resizebox{0.85\hsize}{!}{\includegraphics{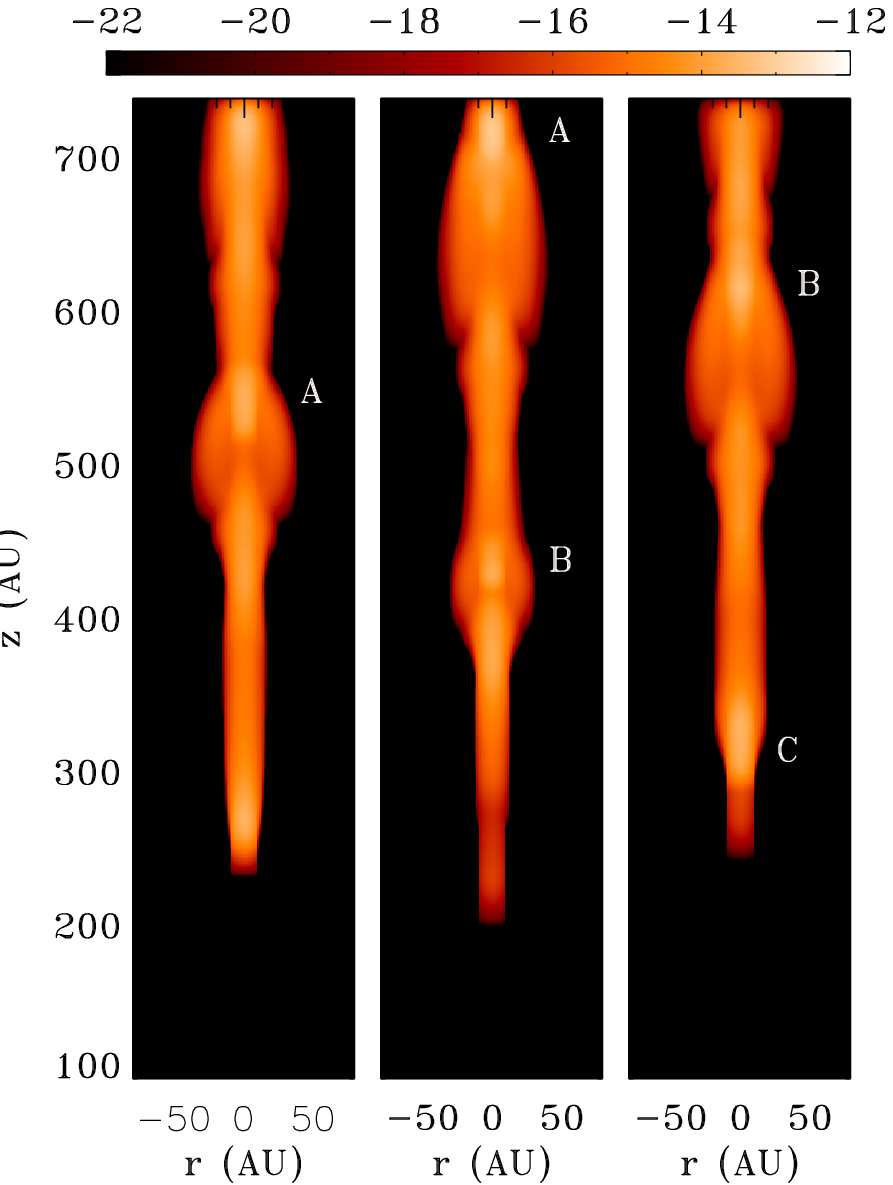}}
  \caption{Logarithmic maps of the surface brightness of the forbidden doublet
    [\ion{S}{ii}], convolved with the PSF, for the three moments shown in
    Fig.~\ref{fig:dn_tmp_mineq}.
    The image is in units of
    $\mathrm{erg}\,\mathrm{cm}^{-2}\,\mathrm{arcsec}^{-2}\,\mathrm{s}^{-1}$ and
    has been projected with a declination angle of $80^\circ$ with respect to
    the line of sight.
    The knots are labeled with letters.}
  \label{fig:emiss_evol_mineq}
\end{figure}
Post-processing was applied on the simulation employing the sophisticated
cooling function MINEq.
In Fig.~\ref{fig:emiss_evol_mineq}, surface brightness maps of [\ion{S}{ii}] are
shown for the corresponding three outputs of Fig.~\ref{fig:dn_tmp_mineq}, having
also applied a declination angle of $80^\circ$.
The outflow appears as a prominent well-collimated jet with a small opening
angle.
In fact, the overall emission seems to originate from the region enclosed by the
weak oblique shock discussed in Sect.~\ref{sec:dynamics}.
For our model, this suggests that the responsible mechanism for heating the bulk
of the flow at the temperatures required for an observable emission is the
effect of recollimation, which compresses the flow around the axis.
The presence of the less dense and slower surrounding wind cannot be seen in
this emission line.
Nevertheless, it fills up the rest of the computational domain and is considered
to emanate from the outer disk radii, contributing to the mass and angular
momentum loss rates.

\begin{figure}
\centering
  \resizebox{\hsize}{!}{\includegraphics{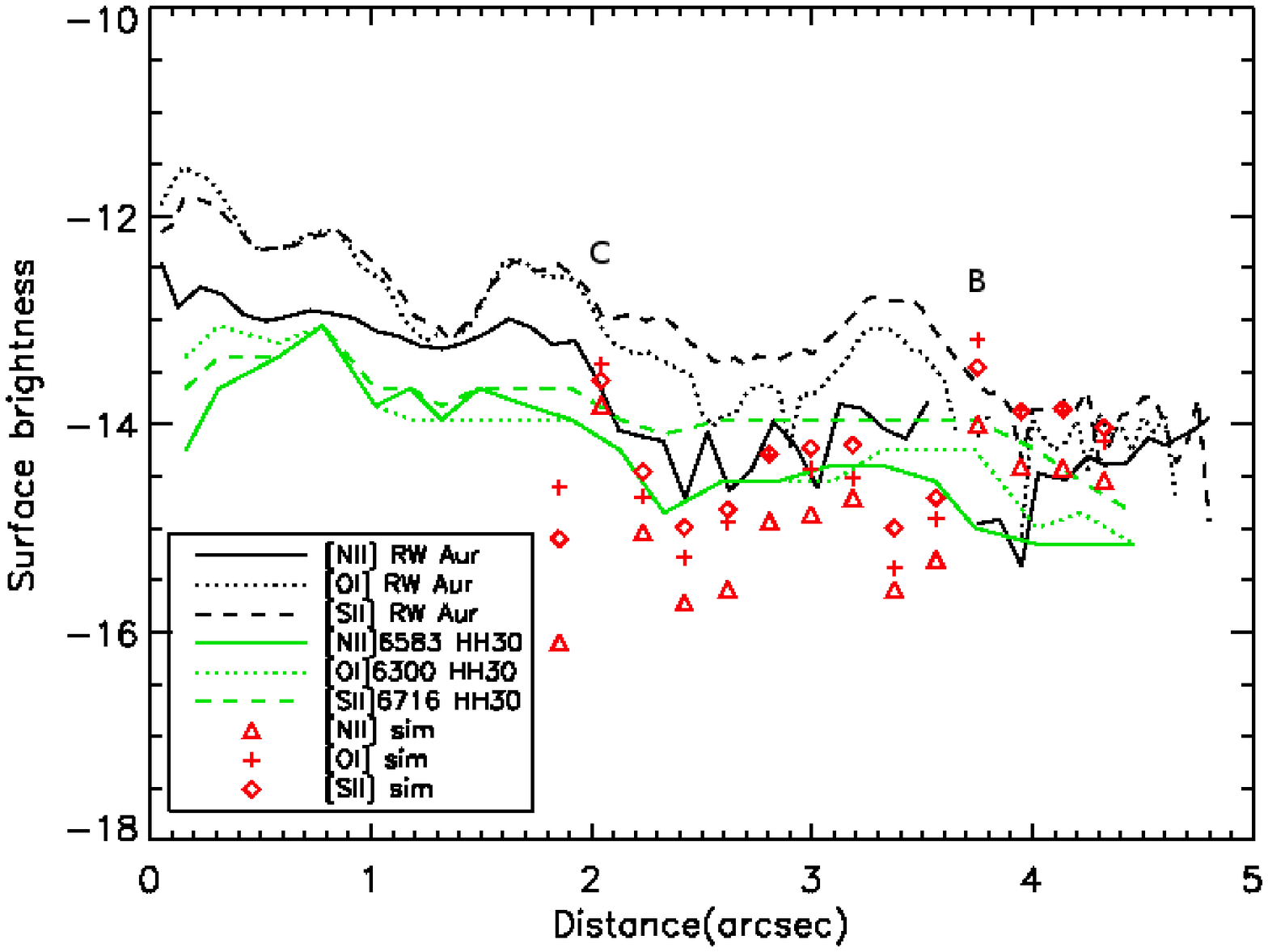}}
  \caption{Logarithmic surface brightness profiles of the forbidden doublets
    along the axis [\ion{S}{ii}], [\ion{N}{ii}], and [\ion{O}{i}], for two
    observed YSO jets (RW Aurigae and HH\,30) and the MINEq simulation.
    The image is in units of 
    $\mathrm{erg}\,\mathrm{cm}^{-2}\,\mathrm{arcsec}^{-2}\,\mathrm{s}^{-1}$ and
    corresponds at $t = 21.6$ yrs.}
  \label{fig:compare_obs}
\end{figure}
Moreover, emission knots are observed along the jet axis, which correspond to
high density and temperature regions produced by the propagating shocks.
The knots have an enhanced surface brightness which can be $10$ times higher
than that of the bulk flow, see Fig.~\ref{fig:compare_obs}.
The introduced variability has been intentionally chosen to be on the order of a
few years, producing knot structures every few hundred $\mathrm{AU}$.
This is the typical length scale observed in YSO jets, as for example in the
systems HH\,1\&2, HH\,34, and HH\,47, for which high angular resolution
astronomical data are available (e.g. Hartigan et al. \cite{Har11}).

Furthermore, Fig.~\ref{fig:compare_obs} compares directly the line emissions
given by the simulation with those of two observed jets, RW Aur and HH\,30.
The plot suggests a good agreement in the intensities at large distances from
the origin of the outflow.
In fact, the emission coming from the bulk flow of this model is closer to
observations than the earlier work of Te\c{s}ileanu et al. (\cite{Tes12}).
The improvement is attributed to the heating provided by the recollimation of
the flow.
Moreover, a pre-existing ionization would provide a higher emissivity in the
regions between the knots bringing the model closer to observations.

The discrepancy close to the source might be evidence for additional mechanisms
present in the first part of the jet propagation, such as a pre-existing heating
and ionization of the gas that can increase emissivity (e.g. Te\c{s}ileanu et
al. \cite{Tes12}).
On the other hand, the flow has to propagate for a certain distance before the
induced sinusoidal time dependence can form shocks.
In addition, the jet is initially expanding before it recollimates and heat the
gas.
We note that low emission close to the central object is also found
observationally, for instance in the jets HH\,34, HH\,211, and HH\,212 (e.g.
Correira et al. \cite{Cor09}).
However, the length of the low emissivity region of those objects is at least by
an order of magnitude larger than here, and also the corresponding mechanisms
and/or the parameters used in the present simulation, maybe different.

\begin{figure}
\centering
  \resizebox{0.85\hsize}{!}{\includegraphics{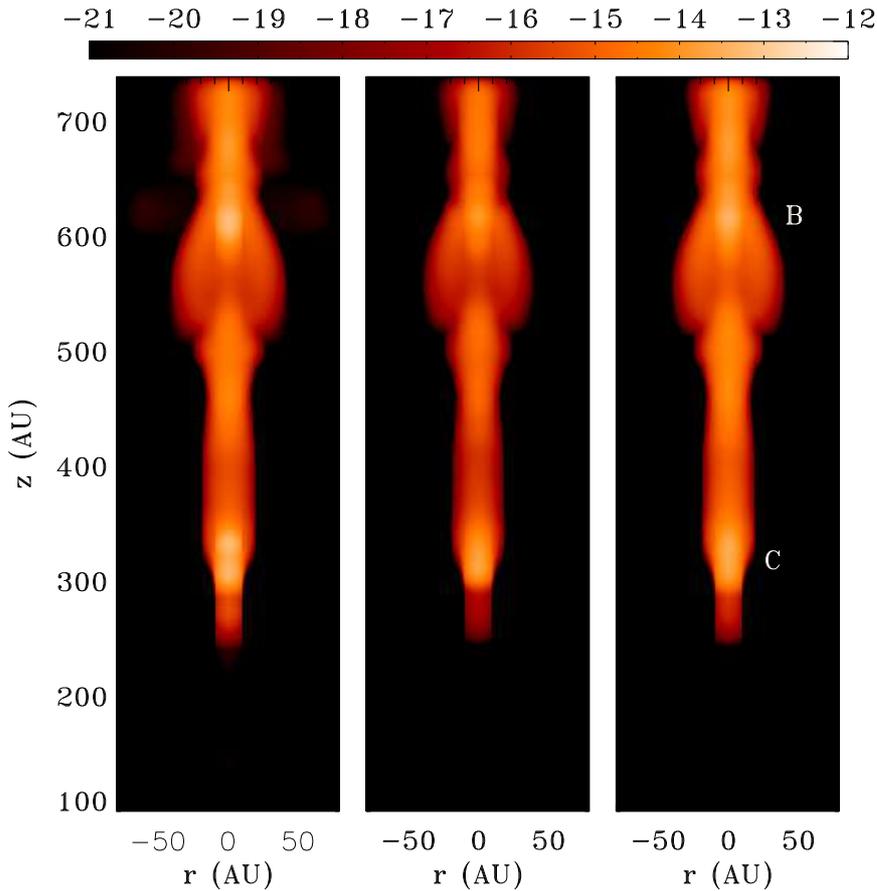}}
  \caption{Logarithmic maps of the surface brightness, processed as described in
    Fig.~\ref{fig:emiss_evol_mineq}, for the three forbidden doublets of
    [\ion{O}{i}], [\ion{N}{ii}], [\ion{S}{ii}] (from left to right), in units of
    $\mathrm{erg}\,\mathrm{cm}^{-2}\,\mathrm{arcsec}^{-2}\,\mathrm{s}^{-1}$.
    The snapshot corresponds to $t = 21.6\,\mathrm{yrs}$.}
  \label{fig:emiss_mineq}
\end{figure}
Finally, Fig.~\ref{fig:emiss_mineq} displays the surface brightness maps for the
three forbidden doublets of [\ion{O}{i}], [\ion{N}{ii}], and [\ion{S}{ii}] for
the last evolutionary moment displayed in Fig.~\ref{fig:dn_tmp_mineq}.
All three panels highlight the same structure.
The two knots, located slightly above $\sim$$300\,\mathrm{AU}$, and
$\sim$$600\,\mathrm{AU}$, emit more strongly than the rest of the flow and
can be clearly seen in all three emission lines.
We note that we have not simulated the bow shock where the outflow interacts
with the interstellar medium, since our initial conditions did not include it.

%%%%%%%%%%%%%%%%%%%%%%%%%%%%%%%%%%%%%%%%%%%%%%%%%%%%%%%%%%%%%%%%%%%%%%%%%%%%%
\subsection{PV diagram}

\begin{figure}
\centering
  \resizebox{\hsize}{!}{\includegraphics{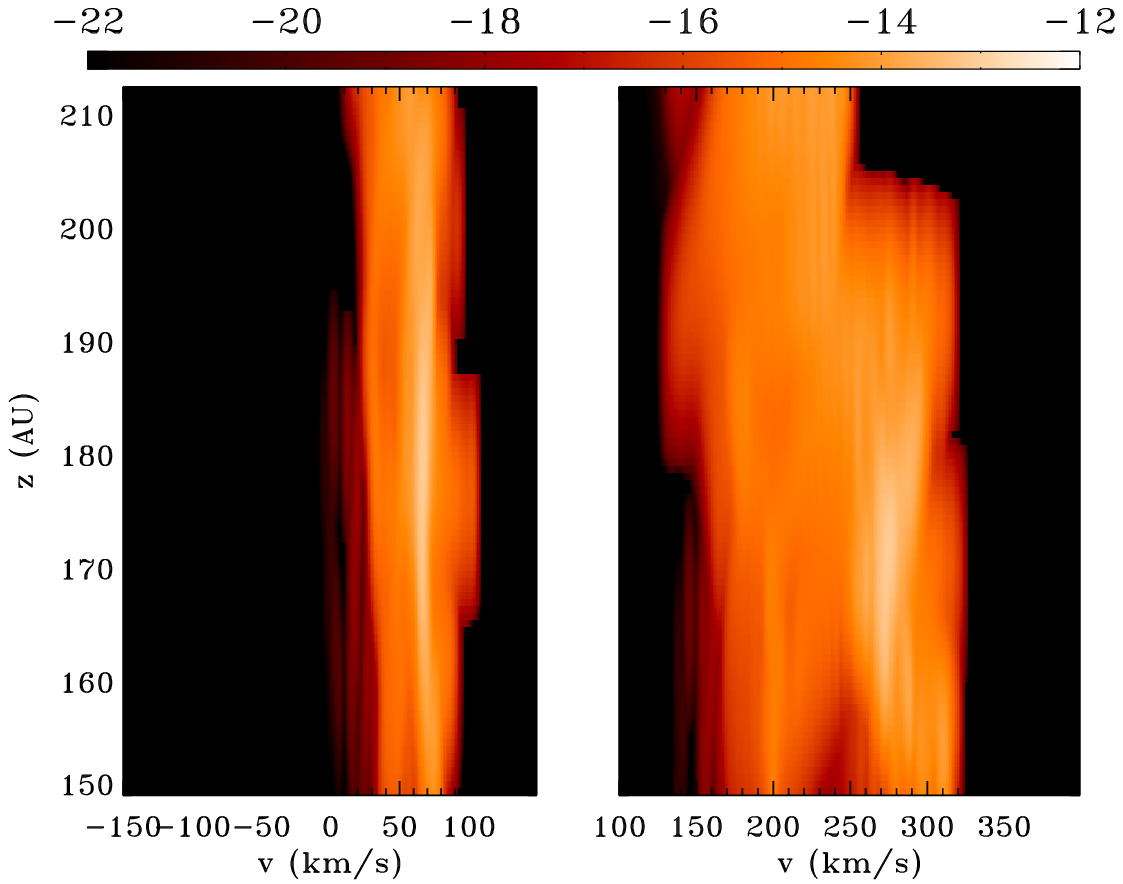}}
  \caption{Position-velocity diagrams, for a declination angle with the line of
    sight of $80^\circ$ (left) and $45^\circ$ (right), for the forbidden doublet
    of [\ion{S}{ii}].
    Units in
    $\mathrm{erg}\,\mathrm{cm}^{-2}\,\mathrm{arcsec}^{-2}\,\mathrm{s}^{-1}$.}
  \label{fig:pv_mineq}
\end{figure}
Figure~\ref{fig:pv_mineq} displays examples of the Position-Velocity (PV) 
contours, a diagram that is widely used for the representation of observational 
data of YSO jets.
It shows the distribution of the brightness with respect to the position along
the spectrometer slit and the velocity along the line of sight.
We have chosen two declination angles for the reconstructed 3D distribution and
an arbitrary slit width of $20\,\mathrm{AU}$ positioned along the central jet
axis.
In the case of a jet almost perpendicular to the line of sight ($80^\circ$,
left panel), the projection of the longitudinal speed is small and hence a value
of $\sim$$60\,\mathrm{km\,s}^{-1}$ is recovered at all heights, with deviations
on the order of a few tenths of $\mathrm{km\,s}^{-1}$.
However, when the angle between the line of sight and the outflow axis is
smaller ($45^\circ$, right panel) the projected component is larger, around
$\sim$$250\,\mathrm{km\,s}^{-1}$.
The distribution of velocities is also wider,
$\sim$$50\,\mathrm{km\,s}^{-1}$, since the flow fluctuations can be now
observed.
In addition, the figure makes evident the parts of the flow that propagate
faster and hence the speed distribution can be inferred and associated with the
observed knot structures.

For the latter case of a lower declination angle with respect to the line of
sight, the surface brightness maps in the emission lines of interest are similar
to Fig.~\ref{fig:emiss_mineq}, apart from the distance between the knots which
seems shorter due to projection effects.

%%%%%%%%%%%%%%%%%%%%%%%%%%%%%%%%%%%%%%%%%%%%%%%%%%%%%%%%%%%%%%%%%%%%%%%%%%%%%
\subsection{Jet radius}
  \label{sec:jetradius}

We proceed to calculate the jet radius from Fig.~\ref{fig:emiss_mineq}.
We follow a simple approach based on the Full Width Half Maximum (FWHM) method.
The maximum at each height of the emission map is calculated and then the radius
where the distribution takes half that value is determined.
Since the plot is logarithmic, the background emission is required in order to
correctly compute the height of the distribution over the radius.
We have considered this parameter to be $10^{-26}\,
\mathrm{erg}\,\mathrm{cm}^{-2}\,\mathrm{arcsec}^{-2}\,\mathrm{s}^{-1}$, but we
note that values different by a few orders of magnitude provide almost the same
result.

\begin{figure}
\centering
  \resizebox{\hsize}{!}{\includegraphics{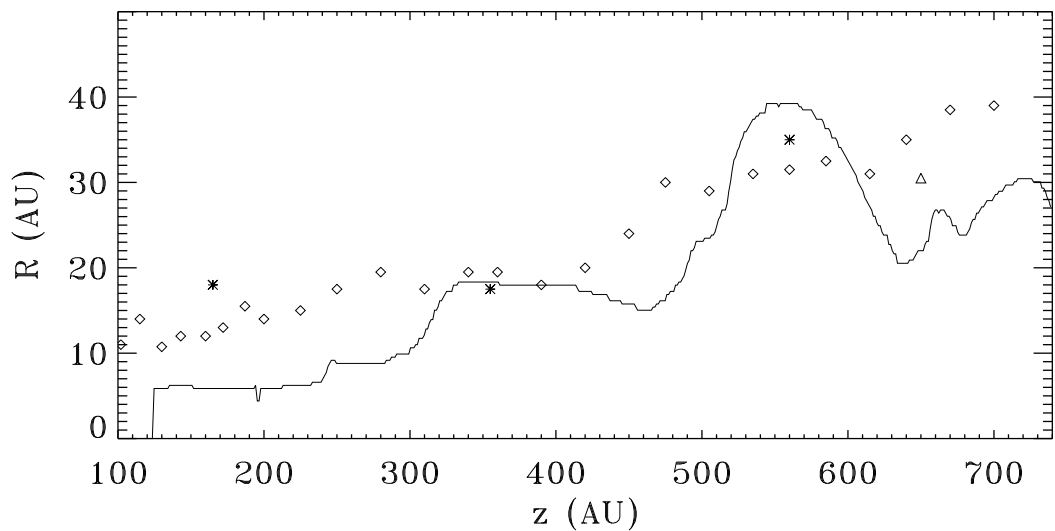}}
  \caption{The jet radius (solid line), $R$, as computed from the average of the
    doublets \ion{O}{i}, \ion{N}{ii}, and \ion{S}{ii} for $t = 21.6$.
    Diamonds denote observations of RW Aur, stars of HH\,30, and triangles of HL
    Tau.}
  \label{fig:radius_mineq}
\end{figure}
Figure~\ref{fig:radius_mineq} plots the jet radius, $R$, as computed from the
average of the doublets \ion{O}{i}, \ion{N}{ii}, and \ion{S}{ii}.
Data points of the jets RW Aur, HH\,30, and HL are also shown for comparison.
The plot suggests a jet width on the order of $40$ to $60\,\mathrm{AU}$, in good
agreement with observations (e.g. Ray et al. \cite{Ray07}, and references
therein).
The variations of $R$ along the axis is due to the applied speed variability.
However, this does not seem to disrupt the average width, even though it
introduces local deviations.
Apart from the jet radius, also the opening angle is comparable and is of 
the order of a few degrees.

Stute et al. (\cite{Stu10}) found similar jet radii for their truncated
disk-wind solution.
The radii for the untruncated cases were much larger compared to the present
study, mostly because of the absence of the cooling term in the energy equation
which resulted in much higher temperatures.

%%%%%%%%%%%%%%%%%%%%%%%%%%%%%%%%%%%%%%%%%%%%%%%%%%%%%%%%%%%%%%%%%%%%%%%%%%%%%
%
\section{Summary -- Conclusions}
  \label{sec:conclusions}
%
%
%%%%%%%%%%%%%%%%%%%%%%%%%%%%%%%%%%%%%%%%%%%%%%%%%%%%%%%%%%%%%%%%%%%%%%%%%%%%%

In this paper, we started from a combination of two analytical outflow
solutions, a stellar jet and a disk wind, we simulated the two-component jet,
and we then generated synthetic emission maps.
We carried out 2.5D axisymmetric numerical simulations adopting a sophisticated
cooling function that follows the ionization and optically thin radiation losses
of several ions.
We also applied a velocity variability at the base of the outflow in order to
produce shocks and knots along the axis.

Our conclusions can be summarized as follows:
\begin{enumerate}
\item
The dynamical evolution of the two-component jet model is similar whether a
simplified or a detailed cooling function is adopted.
However, the adiabatic case leads to the overestimation of the jet radius by a
factor of $2$ and the temperature by an order of magnitude.
\item
The density, temperature, and velocity along the axis are within the typical
value range of observed astronomical sources.
\item
Apart from the above physical parameters, the jet radius as well as the opening
angle are also found to be close to typical YSO jets.
\item
The dense and hot inner part of the jet emits strongly, with the synthetic
emission maps showing a well-collimated outflow that resembles closely real
observations.
The emission knots propagate along the axis and demonstrate enhanced emission
being dense post-shock regions.
\item
The predicted emission lines match the observations of YSO jets at high
altitudes, but they have smaller values close to the source.
We speculate that by taking into account heating and pre-ionization at the base
of the flow might reduce this discrepancy.
\end{enumerate}

Our results are very encouraging and prompt for further investigation.
By closing the gap between analytical solutions, numerical simulations and 
observations, a valuable feedback is provided that will help to further improve
the outflow models as well as understand deeper the jet phenomenology.
The simulations reported here are axisymmetric and may be expanded to allow
nonaxisymmetric perturbations.
However, note that the recent 3D simulations of disk-winds crossing the FMSS
have shown that the analytical MHD solutions behave well even when the basic
assumption of axisymmetry is relaxed (Stute et al. submitted).

In a future work, we plan to model specific YSO jets in an attempt to recover
their properties and understand their dynamics. Further work will also focus on
the introduction of pre-existing ionization of the jet material, likely to 
improve the agreement of the simulation results with observations. From the
synthetical observations point of view, the present results seem to be 
complemented by the ones previously published in Te\c{s}ileanu et al.
(\cite{Tes12}) for the first part of jet propagation (the first 2 arcseconds).

\begin{acknowledgements}
  We are grateful to an anonymous referee whose constructive report resulted in
  improving and clarifying several parts of the paper.
  This research was supported by a Marie Curie European Reintegration Grant
  within the 7th European Community Framework Programme, (``TJ-CompTON'',
  PERG05-GA-2009-249164), and also partly by the ANR STARSHOCK project
  ANR-08-BLAN-0263-07-2009/2013.
  The numerical simulations were performed on the cluster at the Laboratory 
  of Computational Astrophysics at the Faculty of Physics, University of
  Bucharest. The contribution of OT is based on the research performed during 
  the project financed by contract CNCSIS-RP no. 4/1.07.2009 (2009-2011) in 
  the Romanian PNII framework. 
\end{acknowledgements}

\end{document}